\documentclass[epj,final]{svjour}
\usepackage{latexsym,psfrag}
\usepackage{graphics}
\usepackage{psfig,epsfig}
\usepackage{graphicx,amsmath,amssymb}
\newcommand{\p}[1]{\left({#1}\right)}
\newcommand{\pq}[1]{\left[{#1}\right]}

\begin{document}
\title{Kinetic barriers in RNA unzipping}
\author{Alberto Imparato\thanks{To whom correspondence should be addressed. E-mail:
\texttt{imparato@na.infn.it}} \and Luca Peliti\thanks{Associato
INFN, Sezione di Napoli.}} \institute{Dipartimento di Scienze
Fisiche and Unit\`a INFM,
Universit\`a ``Federico II''\\
Complesso Universitario di Monte S. Angelo, I--80126 Napoli
(Italy)}
\date{Received: date / Revised version: date}

\abstract{We consider a simple model for the unfolding of RNA tertiary structure under dynamic loading.
The opening of such a structure is regarded as a two step process,
each corresponding to the overcoming of a single energy barrier.
The resulting two-barrier energy landscape accounts for the dependence of the unfolding kinetics
on the pulling rate.
Furthermore at intermediate force, the two barriers cannot be distinguished
by the analysis of the opening kinetic, which turns out to be dominated
by a single macro-barrier, whose properties depend non-trivially on the two single
barriers.
Our results suggest that in pulling experiments on RNA molecule containing tertiary structures,
the details of the single kinetic
barriers can only be obtained using a low pulling rate value, or in the high
force regime.
\PACS{
      {87.14.Gg}{DNA, RNA}   \and
      {82.37.-j}{Single molecule kinetics}
     }
}
\maketitle
\section{Introduction}
The study of structural properties of biological molecules has
received a boost by the introduction of techniques allowing for
the manipulation of single molecules. For example, the study of
folding and unfolding of nucleic acids can now be performed by
applying a controlled force on the free end of a single strain of
a molecule. In this situation, the opening of the Watson-Crick
pairs leads to what has been called the \emph{unzipping} of the
molecule. Unzipping in DNA is similar to some steps involved in
DNA replication and in its translation into mRNA, and has been the
subject of several theoretical and experimental
investigations~\cite{Danilowicz,LN}. On the other hand, RNA unzipping
exhibits further complications, since a given single stranded RNA
molecule can exhibit a complex secondary structure (matching
pattern between complementary bases) and tertiary structure
(three-dimensional conformation). Thus, single-molecule unzipping
experiment can yield information on the secondary structure of RNA
molecules \cite{bus1,bus2,isamb,GBH,MKM}. 
In ref. \cite{isamb} in particular, the role of the secondary structure intermediates
in the folding/unfolding experiments is discussed, and it is shown that 
such intermediates can be responsible for the slowing down of the kinetics.
Moreover, the response of
complex RNA structures to applied mechanical forces can be
analogous to the responses of RNA during translation or export
from the nucleus~\cite{bus2}. However, in such cases, the tertiary
RNA structure plays an important role. It is commonly believed
that the breaking of RNA tertiary structures  or their formation are the
time limiting processes in the molecule unfolding or refolding,
respectively \cite{bus1,SosPan}. In the unzipping experiments,
tertiary contacts may lead to the appearance of kinetic barriers
where the unzipping process momentarily stops \cite{bus1,bus2}. In this situation,
overcoming these barriers bears some analogy with the breakup of
molecular adhesion studied in a number of
experiments~\cite{ev1,ev2} which have been recently reanalyzed
theoretically~\cite{Str,denis1,denis2}. In the present paper, we
wish to understand some features of the experiments on RNA
unzipping in the light of this theory, in order to highlight what
information one is able to collect from the unzipping kinetics on
the position and the height of the barriers due to the tertiary
structure.
Although in the case of the folding of large complex RNA molecules, the formation of kinetically trapped intermediates in the secondary structure can play an important role in the 
slowing down of the  process \cite{TW1,TW2,Woo}, 
here we will specifically consider the role of the tertiary
structures in the unzipping experiments. Such role was stressed 
in refs. \cite{bus1,bus2}, where
the mechanical unfolding of  RNA molecules  were performed. In those
experiments the tertiary
contacts could be removed by changing the solution the RNA was immersed
in. The removal of the tertiary structures corresponded to the disappearing
of the kinetic barriers, and furthermore the folding/unfolding processes became reversible. This kind of experiments indicate that the tertiary structures
are responsible for the kinetic arrest of the structural rearrangement of RNA molecules under tension,  although the  tertiary structures are much more brittle
compared to the secondary ones, i.e they break  in consequence of  small deformations.

\section{The model}
A typical unfolding experiment consists in holding one molecule's
free end in an optical trap while the other free end is pulled  at
constant velocity \cite{bus1,bus2}. This induces a pulling force
on the molecule that increases linearly with time $f(t)=r t$,
where $r$ is the loading rate. The typical output of such
experiment is a force-extension curve where the monotonic increase
of the force is interrupted by a number of plateaus revealing the
unfolding of portions of the molecule.

The unfolding of the molecule is a stochastic process, which
depends on the pulling force rate $r$, the actual value of the
pulling force and the microscopic details of the molecule.
Therefore the unfolding of an RNA molecule, i.e., the successive
breaking of its molecular links, can be viewed as a succession of
thermally activated escape processes over a set of energy
barriers, each representing one or more molecular
links~\cite{SosPan,TW1,TW2}. Within this picture, the resulting
energy landscape can be considered one-dimensional, since the
experimental set-up singles out a well-defined direction, which is
the pulling direction. The landscape energy barriers are thus
located at increasing distance along the pulling direction. This
experiment can be repeated for several values of $r$, in such a
way that the different force-extension curves can be mapped onto a
single curve of the breaking force as a function of the loading
rate. From this curve, much information on the microscopic details
of the molecular bonds can be obtained~\cite{ev1,ev2}: in
particular the typical length and energy of such bonds. Another
quantity, which can be obtained by the force-extension
characteristics, is the fraction $\phi$ of molecules which remain
folded as a function of the time or the force, and which also
depends on the parameter $r$. Sampling the fraction of folded
molecules at different times provides a more direct insight into
the kinetic barriers which slow down the unfolding process. Using
Evans' results for the case of pulling experiments on a single
molecular link \cite{ev2}, which can be represented by a single
kinetic barriers of height $\Delta E$ and position $\Delta x$
along the reaction coordinate, the fraction $\phi(t)$ of bound
links at time $t$ is given by
\begin{eqnarray}
\phi(t)&=&\exp\left[-\int_0^td  t'\; \omega_0\, e^{-\beta
\left(\Delta E -rt'\Delta x\right)} \right]\nonumber\\
&=&\exp\left[-\frac{\omega_0}{\beta r \Delta x}
e^{-\beta \Delta E} \left(e^{\beta rt\Delta x}-1\right)\right] ,
\label{phit}
\end{eqnarray}
where $\omega_0$ is the attempt frequency, which depends on the
microscopic details of molecular linkage, and
$\beta=1/k_\mathrm{B}T$.

In the large force regime ($f\ge 10 $ pN), one expects that the quantity $q(f)$
defined as
\begin{equation}
q(f)=\left.\ln \left[r \ln \frac{1}{\phi(t)}\right]\right|_{t=f/r} ,
\label{qt}
\end{equation}
will be a linear function of $f$:
\begin{equation}
q(f)\simeq \ln \left(\frac {\omega_0} {\beta \Delta x}\right)-\beta \Delta E +\beta  f \Delta x\, ,
\label{linearq}
\end{equation}
and will exhibit no dependence on $r$.

This expression has been exploited by Liphardt et al.~\cite{bus1}
to characterize the tertiary structure of an RNA molecule: by mechanically
pulling on the P5abc domain
of the \textit{Tetrahymena thermophila} ribozyme, the fraction of folded
molecules as a function of the force (time) has been determined. The authors then make
the hypothesis that the tertiary structure can be described as a single
kinetic barrier, which hinders the molecule from unzipping.
Using eq.~(\ref{linearq}), they give an estimate for the two
characteristic parameters of the barrier, i.e.
the zero-force transition rate, defined as
\begin{equation}
k_0=\omega_0\exp\left(-\beta \Delta E\right)\, .
\label{k0}
\end{equation}
 and the barrier position  along the reaction coordinate $\Delta x$ (in that
 case and in the following in this paper,
the elongation of the molecule needed to break the bond).
However, the data for $q(f)$ showed in that paper, exhibit a dependence on
$r$ even if not distinct, in contrast with eqs.~(\ref{qt})
and (\ref{linearq}).

In the present paper we will argue that at least another kinetic
barrier has to be considered in the energy landscape of the tertiary structure
in order to account for the dependence of $q(f)$ on $r$.
Furthermore, in a recent paper, Bartolo et al.~\cite{denis2} showed that in the more complicated case of several
molecular links, which can be represented by a set of $N$ kinetic barriers along a one-dimensional
unbinding path, the unbinding force plotted as a function of logarithm of the pulling rate ($\ln r$)
appears as a succession of straight lines whose slopes are given by the distances between
the adjacent maxima and the minima of the energy landscape.
Thus, we expect that for the simple two-barrier energy landscape here
considered, the plot of $q$ as a function of $f=rt$ will exhibit more than one
straight line, each corresponding to a different escape route from the folded
to the unfolded state.

\begin{figure}[h]
\center
\psfrag{xa}[ct][ct][.7]{$x_a$}
\psfrag{xA}[ct][ct][.7]{$x_A$}
\psfrag{xB}[ct][ct][.7]{$x_B$}
\psfrag{Ea}[ct][ct][.7]{$E_a$}
\psfrag{EA}[ct][ct][.7]{$E_A$}
\psfrag{EB}[ct][ct][.7]{$E_B$}
\psfrag{E}[ct][ct][1.]{$E$}
\psfrag{x}[ct][ct][1.]{$x$}
\includegraphics[width=8cm]{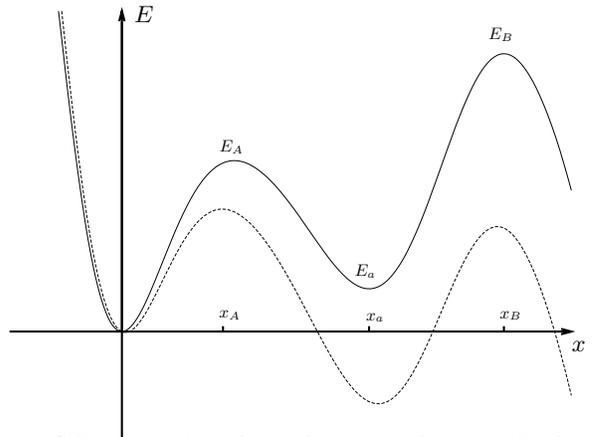}
\caption{Schematic plot of one-dimensional energy landscape with two energy
barriers. The full line corresponds to the landscape with
no external force applied $E(x)$,  while the dotted curve
corresponds to the time-dependent landscape $E(x)-r t  x$ at $t>t_A$
(see text and eq. (\ref{deftA}) for the definition of $t_A$). }
\label{landscape}
\end{figure}
Basically we assume that the tertiary structures of a RNA molecule
can be described by more than one energy barrier with different
width and height: the simplest case is the one depicted in fig.~\ref{landscape}, where a
landscape $E(x)$ characterized by two
barriers is responsible for the slowing of the unfolding process,
and the state $x=0$ corresponds to the unperturbed structure.
The opening of the tertiary structure can thus be regarded
as a two-step process: when the molecule extension is of the
order of $x_A$, as a consequence of thermal denaturation or pulling force,
a first set of tertiary interactions are broken, and
the system jumps into the local energy minimum at $x_a$.
Later, as the length $x_B$ is reached, the remaining tertiary interactions break down.
The choice of a two-barrier landscape has been also suggested to us by the following experimental observation: in RNA pulling experiments, a small but finite 
fraction of molecules under
dynamic loading unfolds with two successive rips \cite{bus1}, revealing the existence of an intermediate state between the completely folded and the unfolded
ones.

When force is applied, the energy landscape changes as $E(x,t)=E(x)-rtx$,
so that the outermost barrier height decreases faster than the inner one: the
unfolding kinetics is thus characterized by two crossover times
which will be respectively indicated with $t_A$ and $t_a$ in the following.
For times longer than $t_A$, the inner barrier becomes the dominant one and
is mainly responsible for the slowing down of the unfolding process with respect to
the $f=0$ case.
Eventually, at times longer than $t_a$ (with $t_a>t_A$), the well $(x_a,E_a)$ disappears
and the unfolding simply results from the overcoming of the first barrier
$(x_A,E_A)$.
If we make the further assumption that the energy
barriers and wells sketched in fig.~\ref{landscape} are sharp, so that the barrier
and well positions remain essentially constant with time, the crossover time $t_A$ is given by
\begin{equation}
t_A=\frac{E_B-E_A}{r(x_B-x_A)}\, ,
\label{deftA}
\end{equation}
and $t_a$ is given by
\begin{equation}
t_a=\frac{E_B-E_a}{r(x_B-x_a)}\, .
\label{ta}
\end{equation}
The last equation also defines a crossover force $f_a=r\cdot t_a$.
Using a Kramer formalism, and the notations shown in fig.~\ref{landscape},
we can write the instantaneous rate for the
transition from one of the two minima ($x=0,x_a$) over the two corresponding
energy barriers ($x=x_A,x_B$):
\begin{eqnarray}
k_{0\rightarrow a}&=&{\omega_0} \exp \left[ -\beta\left(E_A -f(t) x_A\right) \right] , \label{rate1}\\
k_{a\rightarrow 0}&=&{\omega_0} \exp\left[-\beta\left(E_A-E_a -f(t) (x_A-x_a)\right)\right] \label{rate2},\\
k_{a\rightarrow \infty}&=&{\omega_0} \exp\left[-\beta\left(E_B-E_a -f(t) (x_B-x_a)\right)\right] ,
\label{rate3}
\end{eqnarray}
where $x=\infty$ indicates the completely unfolded state.
We assume furthermore $k_{\infty\rightarrow 0,a}=0$, i.e., once unfolded
the system never folds back, as observed experimentally~\cite{bus1}.
Let $p_0(t)$ and $p_a(t)$ be the probabilities that the system is in
the state 0 or $a$, respectively.
The time evolution of this quantities is described by the following
differential equation system
\begin{eqnarray}
\dot p_0&=& -k_{0\rightarrow a} p_0 + \theta(t_a-t) k_{a\rightarrow 0} p_a\, ,
\label{pt0}\\
\dot p_a&=&\theta(t_a-t) k_{0\rightarrow a} p_0-\theta(t_a-t) (k_{a\rightarrow 0}+k_{a\rightarrow \infty})p_a\nonumber\\
&&\qquad{} - \theta(t-t_a) {\omega_0}\, ,
\label{pta}
\end{eqnarray}
where $\theta(t)$ is the Heaviside step function.
The step functions $\theta$ have been included in eqs. (\ref{pt0}) and (\ref{pta}) in order to
take into account the disappearing of the well $(x_a,E_a)$ from the system
energy landscape at $t=t_a$.
It is worth to note that at such crossover time, a significant fraction
of molecules might be accumulated in the $x=x_a$ state, as a result
of the system evolution at previous times.
Thus we assume that at time $t>t_a$,  the escape rate of the molecules which are still in the state $x=x_a$, is determined by the molecular attempt frequency $\omega_0$ alone.
For a given set of characteristic parameter 
\[{\mathcal{S}}=\left\{{\omega_0},x_A,E_A,x_a,E_a,x_B,E_B\right\}\, , \]
given the initial values $p_0(0)=1$ and $p_a(0)=0$, such a
system can be solved numerically.
As mentioned above, the molecule never folds back, once it has been completely unfolded,
therefore the quantity
\begin{equation}
\phi(t)=p_0(t)+p_a(t)
\end{equation}
defines the probability that the system is \emph{still} folded at time $t$,
either completely in the $x=0$ state or partially in the $x=a$ state.

In order to obtain a reliable estimate of the parameter set
${\mathcal{S}}$, we consider the results of the above cited
experiment: applying eq.~(\ref{linearq}) to the openings of a
simple RNA molecule tertiary structure, a zero-force transition
rate $k_0\simeq 2\times 10^{-4}\, $s$^{-1}$ and a difference
between the unfolded tertiary structure length and its transition
(breaking) length $\Delta x\simeq 1.6\, $nm have been obtained
\cite{bus1}. The zero-force transition rate $k_0=2\times
10^{-4}\,$s$^{-1}$ is related to the attempt frequency
${\omega_0}$ and to the overall energy barrier $\Delta E$ of the
single barrier picture via eq.~(\ref{k0}). In order for the
tertiary structure to be the dominant impedance to the molecule
unfolding, the involved energy barriers have to be  greater than
the well known RNA base pair energies, which are of the order of a
few ${k_\mathrm{B}T}$. Thus, if we suppose that the energy barrier
$\Delta E$ is of the order $\Delta E\simeq 10\, {k_\mathrm{B}T}$,
from the above cited result and from eq.~(\ref{k0}), we obtain
${\omega_0}\simeq 4.4\,$s$^{-1}$. Our estimate for the tertiary structure energy is in agreement with the values shown in \cite{sil}, where combining numerical computations with experimental techniques, the energy of tertiary interactions in a simple RNA molecules was found to range between $6\, k_\mathrm{B}T$ and $13\,  k_\mathrm{B}T$.  On the other hand, a direct measurement of the attempt frequency ${\omega_0}$  
 in folding/unfolding experiments is rather difficult. An
indirect estimate can be obtained by pulling the RNA molecule at
constant force, when the force value is within an interval of a
few pNs around the unfolding force. The molecule then hops back
and forth between the folded and unfolded state with a frequency
between $0.05\,  $s$^{-1}$ and $20\,   $s$^{-1}$ \cite{bus1},
which depends on the actual value of the force. It can be assumed
that, at the unfolding force, the effective energy barrier
vanishes and the hopping rate yields a rough estimation of the
microscopic attempt rate. Thus our estimate for the attempt
frequency $\omega_0$ is in agreement with those experimental
results.

\begin{figure}[h]
\center
\psfrag{p0}[rc][rc][.7]{$p_0(t)$ }
\psfrag{pa}[rc][rc][.7]{$p_a(t)$ }
\psfrag{t}[ct][ct][1.]{$t$ (s)}
\includegraphics[width=8cm]{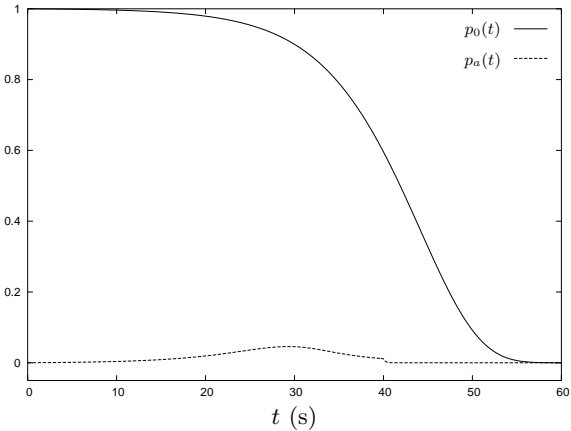}
\caption{Probabilities $p_0$ and $p_a$ as functions of the time, obtained by numerical
integration of eqs. (\ref{pt0}) and (\ref{pta}) with $r=1$ pN/s.}
\label{fpt}
\end{figure}

Here we take $x_A=0.6$ nm,  $x_a=0.8$ nm, and $x_B=1.8$ nm: compared to
 the above cited result, our choice corresponds
 to two successive openings which occur at elongations whose sum
 is equal to the single step picture elongation $\Delta x=1.6$ nm.
The set of parameter values that we will consider in the following is thus
given by
\begin{eqnarray}
\mathcal S &=&\left\{{\omega_0}{=}4.4\, \mathrm{s}^{-1},
x_A{=}0.6\, \mathrm{nm},
E_A{=}10\,{k_\mathrm{B}T},\right.\label{paraset}
\\
&&\left.\quad x_a{=}0.8\, \mathrm{nm},
E_a{=}6\,{k_\mathrm{B}T},
x_B{=}1.8\, \mathrm{nm},
E_B{=}16\,{k_\mathrm{B}T}\right\}, \nonumber
\end{eqnarray}
with $T=300$~K.
Note that with this choice for the energy barrier height,
after the first opening, the system has to overcome another barrier of relative height
$E_B-E_a=E_A$ in order for the second opening to occur.
The two probabilities $p_0(t)$ and $p_a(t)$, obtained by numerical
integration of eqs. (\ref{pt0}) and (\ref{pta}) with $r=1$ pN/s, are shown in fig.~\ref{fpt}.

\section{Results}
\begin{figure}[h]
\center
\psfrag{xA}[rc][rc][.7]{$\ln \left({\omega_0} /\beta x_A\right)-\beta E_A +\beta  f x_A$}
\psfrag{xB}[rc][rc][.7]{$\ln \left({\omega_0} /\beta x_B\right)-\beta E_B +\beta  f x_B$}
\psfrag{e-2}[rc][rc][.7]{$r=10^{-2}$ pN/s}
\psfrag{e-1}[rc][rc][.7]{$r=10^{-1}$ pN/s}
\psfrag{1}[rc][rc][.7]{$r=1$ pN/s}
\psfrag{10}[rc][rc][.7]{$r=10$ pN/s}
\psfrag{100}[rc][rc][.7]{$r=10^2$ pN/s}
\psfrag{1000}[rc][rc][.7]{$r=10^3$ pN/s}
\psfrag{f}[ct][ct][1.]{$f$ (pN)}
\psfrag{q}[cb][cb][1.]{$q$}
\includegraphics[width=8cm]{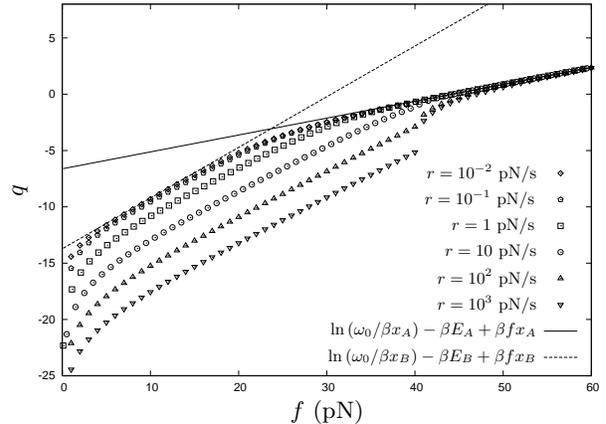}
\caption{Plot of $q$ as a function of $f$, as defined by eq.~(\ref{qt}), for
different values of $r$.
At large force, for each value of the loading rate $r$, the curves all
collapse on a single scaling function
which corresponds to the escape from the innermost barrier $(x_A,E_A)$ (full line).
For small values of $r$, and in the small force regime, the curves also converge
to a function which corresponds to the escape from the outermost barrier $(x_B,E_B)$ (dashed line): in this limit
the structure of the inner part of the energy landscape does not
affect the behavior of the function $q(f)$.}
\label{figureqf}
\end{figure}

In figure~\ref{figureqf} the function $q$, as defined by
eq.~(\ref{qt}), is plotted as a function of $f$ for the energy
landscape represented in figure~\ref{landscape}, with the set of
parameter values ${\mathcal{S}}$ given by eq.~(\ref{paraset}), and
for different values of the loading rate $r$: the behavior of such
function turns out to be dependent on both the force range and the
loading rate. For all the value of $r$ here considered, in the
large force regime $f>f_a$, all the curves collapse on a single
scaling curve, which
 corresponds to the escape from the innermost barrier $E_A$ at $x_A$, i.e., after the outermost
 barrier has disappeared, the innermost barrier
 becomes the only obstacle for the unfolding of the molecule.
On the other hand, for the smallest value of $r$ here considered ($r=10^{-2}$ pN/s),
 and at intermediate force values,  $q(f)$ lies on a line  given by eq.~(\ref{linearq}) with $\Delta x=x_B$ and
 $\Delta E=E_B$.

These results are in agreement with those of the above mentioned work of Bartolo et al.~\cite{denis2},
where the authors found out that slope of the breaking force as a function of
$\ln r$ is  equal to the position of the outermost barrier, in the small
$r$ regime. The same quantity has been found to be equal to  the position of the innermost barrier
at large $r$, if the relative height of the two barriers
is similar, as in our case.
It is worth to remark that in this cited work the single escape rate approximation
has been used, i.e., it has been assumed that the mean escape time from
each of the landscape barrier is constant.
In the present work we do not use such approximation, and thus
our results can be considered more general of those contained in~\cite{denis2}.

\begin{figure}[h]
\center
\psfrag{1}[rc][rc][.7]{$r=1$ pN/s}
\psfrag{10}[rc][rc][.7]{$r=10$ pN/s}
\psfrag{100}[rc][rc][.7]{$r=10^2$ pN/s}
\psfrag{1000}[rc][rc][.7]{$r=10^3$ pN/s}
\psfrag{xm}[rc][rc][.7]{$\ln\left(k(r=1)/\beta x_m\right)+ \beta f x_m$}
\psfrag{f}[ct][ct][1.]{$f$ (pN)}
\psfrag{q}[cb][cb][1.]{$q(f,r)+\ln(r)$}
\includegraphics[width=8cm]{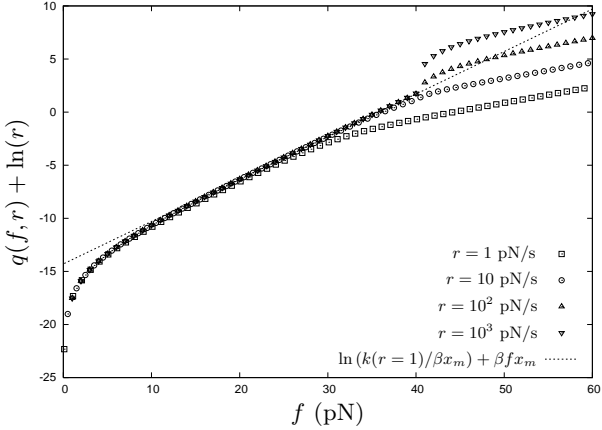}
\caption{Plot of $q(f,r)+\ln(r)$ as a function of $f$. The data all collapse
on the same curve $r=1$, in the force range $ f\lesssim f_a$.}
\label{scaleq}
\end{figure}

\begin{figure}[h]
\center
\psfrag{r}[ct][ct][1.]{$r$ (pN/s)}
\psfrag{p0}[cb][cb][1.]{$p_0(f_a)$}
\includegraphics[width=8cm]{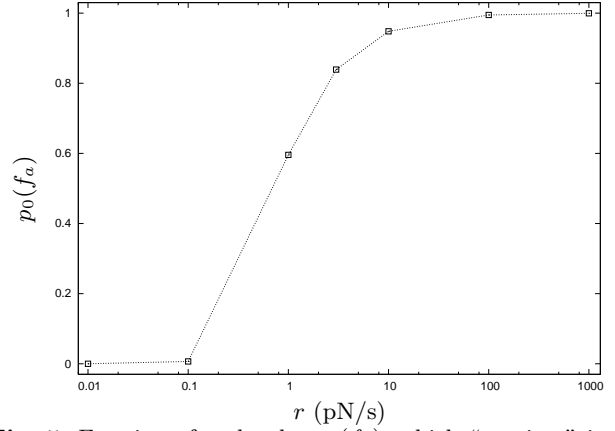}
\caption{Fraction of molecules $p_0(f_a)$, which ``survives'' in the state $x=0$ at the crossover force $f_a$, as
a function of $r$.
The dotted line is a guide to the eye.}
\label{fraction}
\end{figure}

The plot in fig. \ref{figureqf}, indicates that the behavior of $q(f)$, in the
intermediate force range, strongly depends on the value of the pulling rate $r$.
By using  linear fits, we find that for all the curves with $r\ge1$~pN/s, in the force range
$10\, {\rm pN}\lesssim f \lesssim f_a\simeq 40\, {\rm pN}$, the slope
of $q(f)$ is equal to $x_m=x_B-x_a+x_A=1.6$~nm. This indicates that the two 
bonds here considered, behave as a single macro-bond whose typical length
is $x_m$. In analogy with the single bond case,  in
the  force range $10\, {\rm pN}\lesssim f \lesssim f_a$,  the quantity $q$ can be
written as
\begin{equation}
q(f,r)=\ln\left[\frac{k(r)}{\beta x_m}\right]+ \beta f x_m\, ,
\end{equation}
where we have explicitly taken into account the dependence on $r$,  and
where $k(r)$ is the zero-force  transition rate of the macro-bond, which
depends on the details of the energy landscape. Such a quantity decreases
as the loading rate $r$ increases, and it turns out to scale as $k(r)\propto r^{-1}$, as
can be seen in figure \ref{scaleq}.
Using qualitative arguments, Evans \cite{ev2} has proposed that in a set of $N$ identical molecular
bonds in series,  the zero force transition rate decreases as the inverse of $N$.
In our case the inverse proportionality of the  zero-force transition rate on $r$,
appears to depend strongly on the model used here, and in particular on the form
of eqs.~(\ref{pt0}) and (\ref{pta}) used to obtain the quantity $\phi(t)$.
The dependence of $q$
on the loading rate $r$, indicates that at a given force $f\le f_a$, the fraction of molecules
which remain in the state $x=0$ before unfolding completely, increases with the loading rate $r$.
This is also confirmed by a direct plot of the fraction of molecules that are
in the ground state  at the cross-over force $f_a$, as a function
of $r$, see fig.~\ref{fraction}.
In other words as $r$ increases, the molecule is more and more ``frozen'' in its ground
folded state, and only after the outmost barrier disappears ($f=f_a$), the
system unfolds in a way which is determined only by the inner barrier features.

The characteristics of the quantity $q(f,r)$ in the moderate force regime, i.e., the
value of the slope $x_m=x_B-x_a+x_A$ and the scaling law of the zero-force transition rate $k(r)$,
turn out to be fairly universal: we found the same results with different choices of  parameter sets ${\mathcal{S}}$ (data not shown).
This is at variance with the results discussed in \cite{Str,denis2}
 where the slope of the rupture force as a function of $\ln r$, in a given range
 of $r$, has been found to be characterized by a single
length which corresponds to the relative position of the dominant barrier with
respect to the adjacent energy minimum position. In other words,
our results suggest that the unfolding process is
not dominated by a single escape route over a well determinate barrier, but is rather controlled by the interactions
between the two barriers, which determine a different escape route.

\begin{figure}[h]
\center
\psfrag{f}[ct][ct][1.]{$f$ (pN)}
\psfrag{q/pa}[rc][rc][.8]{$\left. \frac{\phi(t,r=10^{-2})}{p_a(t)}\right|_{t=f/r}$}
\psfrag{exp}[rc][rc][.7]{$\exp\pq{\beta\p{E_a-f x_a}}$}
\includegraphics[width=8cm]{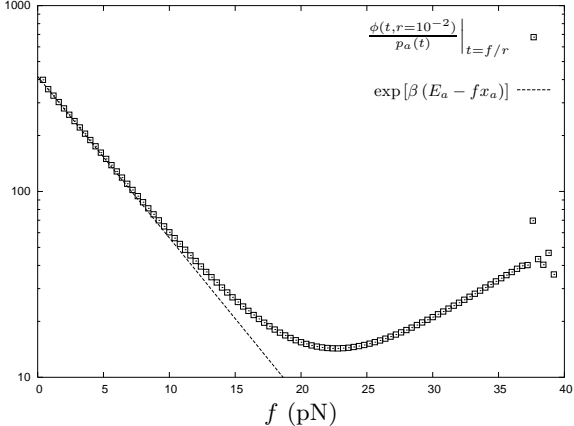}
\caption{Ratio of folded molecule fraction $\phi$ to the fraction $p_a$ of
molecules in the state $x=x_a$, as a function of the force $f$, with $r=10^{-2}$ pN/s.
The dashed line corresponds to the expected function in the small force regime, as given by eq. (\ref{ratio}).}
\label{Ea}
\end{figure}

The analysis of the fraction of folded molecules $\phi(t)$ and its
related function $q(f)$, that we have proposed so far, does not 
provide any estimate of the intermediate energy well $E_a$.
In the following we propose a method to obtain the value of this
quantity, once the fraction of (completely or partially) folded molecules 
$\phi$ and the fraction of molecules which are in the intermediate state $p_a$ have been experimentally determined, as functions of the time (force).
Let us consider the smallest value of the pulling rate we have used
in this paper, $r=10^{-2}$~pN/s. For this value, we have shown that
\begin{equation}
q(f,r=10^{-2})= \ln \p{\frac{\omega_0}{\beta x_B}}-\beta E_B+\beta f x_B\, ,
\end{equation} 
which holds in low-to-moderate force range,
see fig.~\ref{figureqf}.
Using the definition of $q(f)$ as given by eq. (\ref{qt}), we then 
obtain 
\begin{equation}
\phi(t,r=10^{-2})= \exp\pq{-\frac{\omega_0}{\beta r x_B} e^{-\beta\p{E_B-r t x_B}}}\, ,
\label{phi_r_e-2}
\end{equation} 
and
\begin{equation}
\dot \phi(t,r=10^{-2})=-\omega_0 e^{-\beta\p{E_B-r t x_B}} \phi(t,r=10^{-2})\, .
\label{phi1}
\end{equation} 
The last expression is expected to hold for sufficiently small values of 
the force.
On the other hand, summing eq. (\ref{pt0}) and eq. (\ref{pta}), we obtain,
for $t<t_a$,
\begin{equation}
\dot \phi(t)= -k_{a\rightarrow \infty} p_a, 
\label{phi2}
\end{equation} 
where $k_{a\rightarrow \infty}$ is given by eq. (\ref{rate3}).
Putting together eq. (\ref{phi1}) and eq. (\ref{phi2}),
yields
\begin{equation}
\left. \frac{\phi(t,r=10^{-2})}{p_a(t)}\right|_{t=f/r} =e^{\beta\p{E_a-f x_a}}\, .
\label{ratio}
\end{equation} 
Thus we expect that, at low-to-moderate forces where
the equality (\ref{phi_r_e-2}) holds, the ratio of $\phi$ to $p_a$, for small values of the pulling rate, is a linear function
of the force, in a linear-log plot of the data.
This is confirmed by fig. \ref{Ea}, where  the ratio of $\phi/p_a$, as obtained
by numerical solution of eqs. (\ref{pt0}) and (\ref{pta}) with $r=10^{-2}$ pN/s, is plotted as a function of $f$.  From a linear fit of the relative 
experimental  data, it might then be possible to obtain estimates for $x_a$ and $E_a$.

The behaviour of the model here proposed appears to be rather robust with 
respect to  changes in the values of the parameters.

It is clear that, e.g., an increase in $\omega_0$ as well as a (slight) decrease in the barrier heights would yield a corresponding decrease in the time
scales. In the same way, a faster system time evolution could be obtained by, e.g., an increase of either $x_A$ or $x_B$ or a decrease in $x_a$ (see eqs. (\ref{rate1}), (\ref{rate2}) and (\ref{rate3}) which define the transition rates). 

We have shown the numerical results for a energy landscape with $E_A=(E_B-Ea)$.
No qualitative change in the function $q(f,r)$, with respect to that shown in figure \ref{figureqf},  has to be expected by changing
slightly the value of one of the two barriers, while keeping the other one 
unchanged.
Obviously, if one barrier becomes much larger than the other, the behaviour of the single barrier case is recovered.

With our choice for the model parameters, we have implicitly assumed that the inner barrier is the one
which survives longer. Let $t_0$ be the time at which the inner barrier disappears, defined as $t_0=E_A/(r x_A)$, and let $t_{max}$ be the maximum time for the integration of the evolution equations (\ref{pt0}) and (\ref{pta}). 
Thus, with our parameter choice, we took $t_{max}< t_0$. 
This corresponds to stop the  kinetic process before the last barrier disappears: in the absence of barriers the Kramer formalism makes no longer sense.
We now  want to discuss shortly the case where it is the outer barrier to survive longer, i.e. $t_a>t_0$ (and $t_a>t_{max}>t_0$). This can be obtained, e.g., by taking $(E_B-E_a)$ moderately greater than $E_A$.
In this case, a slightly modified version of eqs. (\ref{pt0}) and (\ref{pta}) has to be considered to take into account the disappearing of the barrier $(x_A, E_A)$ at $t=t_0$. The outcome for $q(f,r)$  is similar to that shown in fig. \ref{figureqf}, but the slope at large force is $x_B-x_a$ rather than $x_A$.
Still we believe that the choice $E_A\simeq (E_B-E_a)$, and $x_A\simeq(x_B-x_a)$ is the most reasonable at this level of knowledge of the RNA tertiary structures, since each tertiary contact originates from the same physical mechanisms.
\section{Discussion}
The results shown in figures \ref{figureqf} and \ref{scaleq}
indicate that measuring $q(f)$ at different values of $r$ sheds
light  on different parts of the energy landscape. Direct
information on the outermost barrier can only be obtained by
measuring $q(r,f)$ for very small values of $r$, while for high
values of $f$, one obtains information on the innermost barrier.
The model here presented also provides a method 
to obtain an experimental estimate of the intermediate
energy minimum. 
The measurement of such quantity is highly desirable, since it 
determines the stability of the intermediate state with respect to molecule
pulling.

In the moderate force regime, and for $r\ge 1$ pN/s, the quantity
$q(f,r)$ does not give direct information on a single energy
barrier, but rather indicates that in this force range there is a
strong cooperativity in the unbinding process between the two
kinetic barriers, that can be regarded  as a single one.

In conclusion, we have proposed a simple model for the unfolding of RNA molecules with
tertiary structure under dynamic loading, where two distinct kinetic barriers with similar height, hinder the system
from opening. This choice leads the fraction of folded molecule to depend on
the pulling rate, and therefore accounts for experimental results
where apparently  such a dependence has been observed \cite{bus1}.
In order for our model to be checked, the mechanical openings of a RNA molecule
with tertiary contacts has to be performed with a wide range of pulling
rate. The existence of an unique slope of $q(f)$ vs.\ $f$ in the intermediate
force range, and a scaling law for the zero-force transition rate, as the
one we find out here, would unambiguously indicate that the single barrier
picture is inadequate to describe the kinetic process corresponding to the
mechanical breaking of an RNA tertiary structure.
Our results suggest that the predominance of one of the two single kinetic
barriers on the unfolding process, cannot be inferred by measuring the fraction
of unfolded molecules in a relatively large range of force.
On the contrary, the observation of this quantity over such an intermediate force range,
suggests the existence of a complex kinetic barrier, which has a non trivial
and unexpected connection with the single barriers.
According to our model, in an unfolding experiment, a direct insight into the details of  the single
barriers
can only be obtained either using a relatively small value of the pulling
rate, or analyzing the unfolding kinetics in the large force regime.
However the first possibility is limited by the need to keep the effect of 
 apparatus drift under control.

\begin{acknowledgement}
Useful conversations with A. Ajdari and D. Bartolo, while the authors were
guests of the Laboratoire de Physico-Chimie Théorique de l'ESPCI (Paris), are
acknowledged with pleasure. We are grateful to F. Krzakala for unveiling to
us some of the intricacies of RNA unzipping, and to F. Ritort for observations,  suggestions, and a critical reading of the manuscript.
\end{acknowledgement}

\end{document}